# A general criterion for solid instability and its application to creases

Pengfei Yang[a,*], Yaopeng Fang[a,*], Yanan Yuan[b], Shun Meng[a], Zihao Nan[a], Hui Xu[a], Haroon Imtiaz[a], Bin Liu[a,**], Huajian Gao[c,**]

[a] AML, CNMM, Department of Engineering Mechanics, Tsinghua University, Beijing 100084, China

[b] School of Civil Engineering, Wuhan University, Wuhan 430072, China

[c] School of Engineering, Brown University, Providence, RI 02912, USA (now at College of Engineering, Nanyang Technological University, Singapore)

[*] These authors contributed to the work equally and should be regarded as co-first authors.

[**] Corresponding author.

Email addresses: liubin@tsinghua.edu.cn (B. Liu), huajian.gao@ntu.edu.sg (H. Gao).

**Abstract**: A general force perturbation -based criterion for solid instability is proposed, which can predict instability including crease without priori knowledge of instability configuration. The crease instability is analyzed in detail, we found that the occurrence of solid instability does not always correspond to the non-positive definiteness of global stiffness matrix. An element stiffness-based criterion based on material stiffness is proposed as a stronger criterion in order to fast determine the occurrence of instability. This criterion has been shown to degenerate into the criterion for judging instability of certain known phenomena, such as necking and shear band phenomena. Besides, instability in strongly anisotropic materials is also predicted by the element stiffness-based criterion.

The aim is to develop a more unified theoretical foundation to understand various forms of material instability as well as provide guidelines for the design and fabrication of next generation material systems with built-in instability mechanisms.

1. **Introduction**

Understanding instabilities in solids is important for a large variety of emerging applications, such as buckled ribbons in flexible electronics (Rogers et al., 2010), tunable morphologies in wrinkled surfaces (Zeng et al., 2017; Zong et al., 2016), pattern transformations in phononic metamaterials (Mullin et al., 2007; Rudykh and Boyce, 2014; Wang et al., 2013), and crease-induced fatigue in soft robotics (Hao et al., 2018; Trivedi et al., 2008). Despite the broad interest in the subject, some intriguing instability phenomena, such as the formation of creases in soft materials (Fig. 1d), are still understood at a cursory level.

Previous studies on solid instability have often been conducted via case-by-case analysis of specific phenomena. Relevant literature goes back to as far as Considère (1885), in which the maximum load associated with the onset of necking in a tensile bar was investigated from an experimental perspective. Hill (1962) and Rice (1976) proposed an acoustic-tensor-based criterion for shear band formation in a compressed solid. Although numerous elegant studies have been conducted on specific cases of instability (Bazant and Belytschko, 1985; De Borst et al., 1993; Hadamard, 1903; Santisi d'Avila et al., 2016; Triantafyllidis and Aifantis, 1986), a general criterion that integrates various seemingly different phenomena could foster further understanding and explorations in this field.

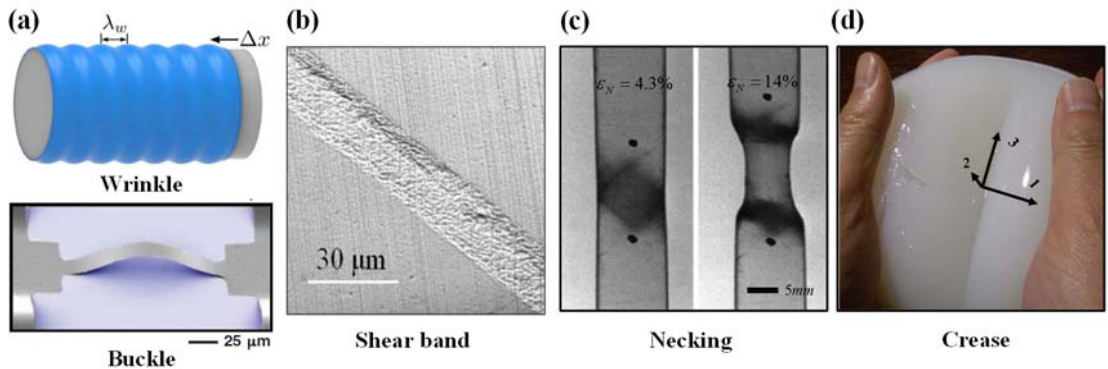

Fig. 1. Various instability phenomena. (a) Structural instabilities: wrinkled shells on substrates (Yang et al., 2018) and buckled ribbons in flexible electronic devices (Rogers et al., 2010); (b) shear band in compressed nanocrystalline Fe (Wei et al., 2002); (c) necking of polyethylene terephthalate (PET) under tension (G'Sell et al., 2002); and (d) crease in a compressed starch gel (Hong et al., 2009).

## 2. A general criterion for solid instability

Stability implies the resistance of a system against perturbation. A general criterion for solid instability in terms of energy perturbation can be given as follows.

*Energy perturbation -based criterion for solid instability*（**criterion 0**）:

**A solid system is in unstable state when it deviates from its current state upon application of an infinitesimal energy perturbation and does not return to its original state after removal of the perturbation. Otherwise, the solid system is in stable.**

Criterion 0 has clear physical meaning therefore, it can be employed theoretically to investigate various instability phenomena. However, the direct application of this criterion through FEM analysis is a challenging task. Therefore, we propose to employ an infinitesimal perturbation force on the solid system in order to realize the infinitesimal energy perturbation. To determine the solid instability through force perturbation, we have the following criterion.

*Force perturbation -based criterion for solid instability*（criterion 1）：

**A solid system is in unstable state, when the application of an infinitesimal perturbation force $\delta \mathbf{f}$ (not necessarily a concentrated force) at equilibrium state leads to a finite perturbation displacement $\delta \mathbf{u}$.**

**Remark1**：It is important to note that Criterion 1 is equivalent to Criterion 0, but it easily be applied for various instability problems. Moreover, Criterion 1 does not require any prior knowledge about the instability configuration, which also shows its applicability for stability analysis.

<u>*Crease analysis through force-based instability criterion*</u>

To demonstrate the validity of criterion 1, we consider a challenging instability problem of crease in a soft material. Crease is a sharp self-contacting fold, which is developed on the free surface of a soft material at a critical strain. Crease has attracted tremendous interest from different scientific communities because it emerges as physical phenomena in diverse range of problems such as crease-induced fatigue in soft robotics (Hao et al., 2018), sulci in biological tissues (Ciarletta et al., 2014; Dervaux et al., 2011; Welker, 1990) and tunable devices (Chen et al., 2014; Kim et al., 2010; Wang et al., 2011). Biot (1963) developed a theoretical formulation for instability in soft material and predicted that the critical strain for surface wrinkling is $-45.6\%$. It is interesting to note that the predicted critical strain remained unchallenged for more than thirty years. Later, Gent and Cho experimentally revealed that the instability in soft material occurs due to crease rather than the wrinkles. Furthermore, the critical strain due to crease is $-35\%$ which is lower than the critical strain for wrinkling (Gent and Cho, 1999). Later, a large number of experimental and numerical studies validated that the critical strain for surface instability is approximately $-35\%$ (Ghatak and Das, 2007; Hohlfeld, 2013; Hong et al., 2009; Tang et al., 2017; Wong et al., 2010). It is important to note that crease and wrinkle are two distinct bifurcation modes (Hohlfeld and Mahadevan, 2011, 2012), but the crease nucleation has not been investigated comprehensively yet. This is due to the fact that the initial assumption of post-instability configurations in previous studies encounters many challenges pertinent to the geometric singularity and the self-contact of a crease (Cao and Hutchinson, 2012; Ciarletta, 2018; Diab et al., 2013; Fu and Ciarletta, 2015; Mora et al., 2011).

In the following, we employ criterion 1 and predict the crease instability without using any assumption of instability configuration.

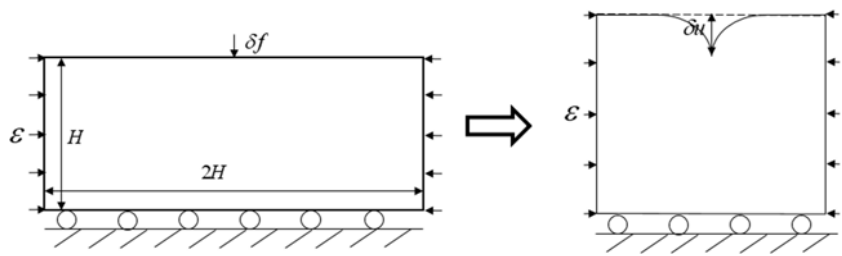

Fig.2. Plane-Strain FEM model

We first consider a material with neo-hooken energy density function:

$$\varphi(\mathbf{C}) = C_{10}\left(\bar{I}_1 - 3\right) + \frac{1}{D}(J-1)^2 \tag{1}$$

where $\mathbf{C} = \mathbf{F}^T\mathbf{F}$ is the right Cauchy-Green deformation tensor, $J = \det(\mathbf{F})$ is the Jacobian of deformation tensor and $I_1$ is the trace of $\mathbf{C}$, $\bar{I}_1 = J^{-\frac{2}{3}}I_1$. Note that the incompressibility limit results in zero value of $D$.

A solid with height $H$ and length $2H$ is subjected to horizontal compressive strain $\varepsilon$ under the plane strain condition and is shown in Fig. 2. The figure shows that the perturbation of concentrated force $\delta f$ is applied to the upper surface. In this study, we employ finite element software ABAQUS6.14 and compute the perturbation displacement $\delta u$ by using with quadrilateral linear element CPE4RH. In order to capture the instability phenomena, we employ a high-density mesh of element size $d$ near the perturbation region as shown in Fig.3a. Figure 3b shows the perturbation force $\delta f$ versus the mesh size $d$ for a given perturbation displacement $\delta u = 0.02H$. It can be seen from Fig. 3b that the perturbation force decreases with the decrease in element size and becomes constant when the element size is $d/H < 0.004$. Since finite element model is more rigid than the real solid material, the larger elements can overestimate the perturbation force. It is important to note that the instability is sensitive to the mode of nonlinear deformation, therefore the dense mesh is recommended to capture the instability phenomenon completely. Although some researchers claimed the existence wrinkle instability instead of crease instability for small perturbation (Wong et al. 2010), but the low-density mesh may cause some inaccuracies in capturing the instability phenomena.

We first obtain the convergence of perturbation force and identify the mesh density for stability analysis. We then compute the compliance $S(\varepsilon)$ by using the $S(\varepsilon) = \lim_{\delta f \to 0} \dfrac{\delta u}{\delta f}$. Figure 4a shows the effect of change in perturbation force on perturbation displacement for several strain cases. We compute the compliance for these strain cases and plot them in Fig. 4b. It can be seen from Fig. 4b that the perturbation compliance approaches to infinity when the compression strain becomes equal to $35\%$. These findings through FEM analysis for plane strain condition are in good agreement with previous studies. Figure 4c shows the crease and the upper surface self-contact phenomena in the deformed configuration at critical compression strain. On the other hand, Fig. 4d demonstrates the non-existence of instability in the deformed configuration at zero compression strain.

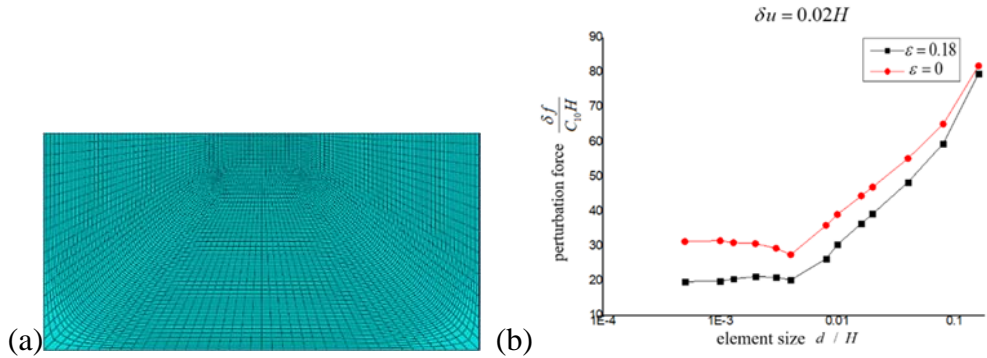

Fig.3. (a) Finite element mesh. (b) Force-element size curve under plane-strain compression.

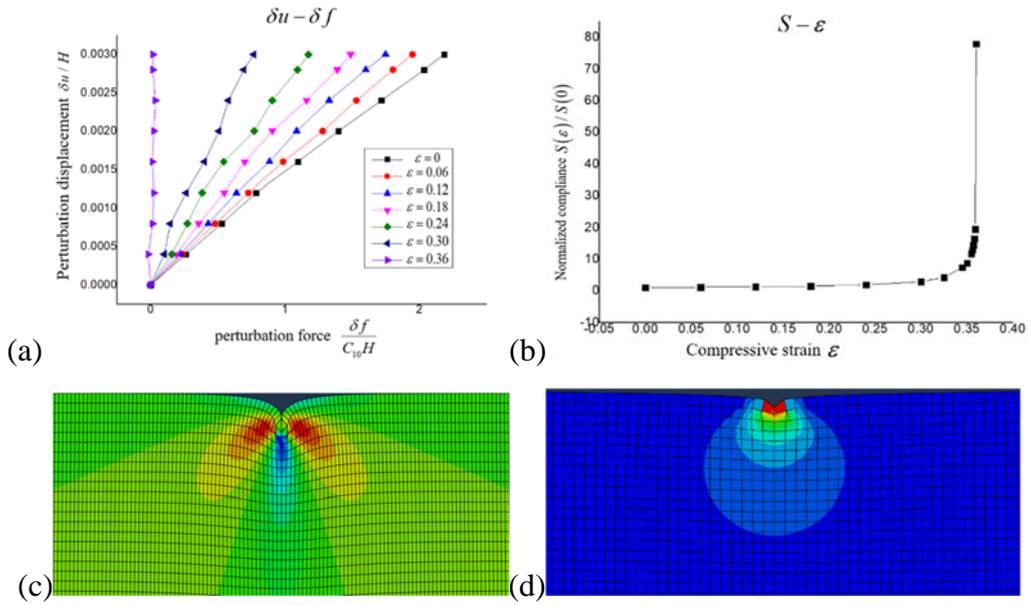

Fig. 4. (a) Force-displacement curve under several plane-strain compression. (b) Compliance-strain curve under plane-strain compression. (c) Post-instability configuration for $\varepsilon = -36\%$. (d) Deformed configuration for $\varepsilon = 0$

**Remark 2: The occurrence of solid instability does not always correspond to the non-positive definiteness of global stiffness matrix.**

To demonstrate the validity of remark 2, we remove the perturbation force and simulate the same problem by using the compression strain only. Here we employ neo-hooken energy density function for solid material and investigate various mesh density cases. It is interesting to note that the application of 40% strain even for dense mesh does not generate any crease in solid material. Figure 5 shows the deformed configurations with 40% strain for several mesh density cases. All cases are deformed uniformly and we did not observe the crease phenomena. Furthermore, the corresponding global stiffness matrix within this compression strain range is always positive definite. The non-positive definiteness of global stiffness matrix was explicitly or implicitly adopted as stability criteria in previous simulations or theoretical analyses (Biot, 1963). However, the present results shows that the non-positive definiteness of global stiffness matrix is not a reliable criterion to identify instability,

A global stiffness matrix is obtained by smooth perturbation displacement field, which implicitly requires $|\delta \mathbf{u}| \ll 1$, $\left|\dfrac{\partial \delta \mathbf{u}}{\partial \mathbf{x}}\right| \ll 1$. However, the material point near the sharp folding region has a large rotation so $\left|\dfrac{\partial \delta \mathbf{u}}{\partial \mathbf{x}}\right| \ll 1$ is not satisfied for crease phenomena. Therefore, the perturbed displacement field for crease cannot be reflected in the searching space of a traditional global stiffness matrix. However, the proposed force-perturbation- -based criterion has no limitation on the magnitude of $\left|\dfrac{\partial \delta \mathbf{u}}{\partial \mathbf{x}}\right|$, so it has wider applicability.

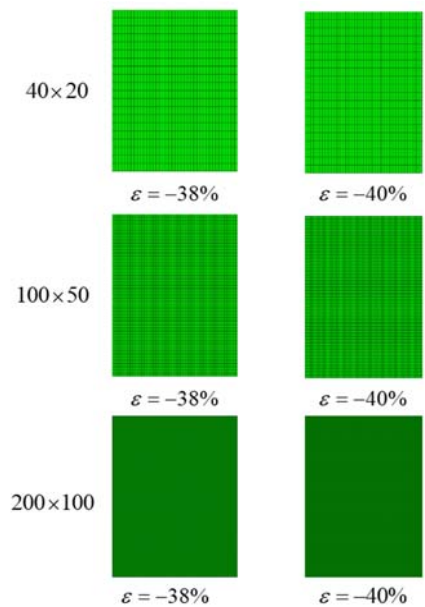

Fig.5. Deformation configurations under different mesh densities

**Remark3**: When the global stiffness matrix is semi-positive definite, a group of infinitesimal perturbation forces can be found to lead to a finite displacement. Therefore, the global-stiffness-matrix criterion is a special case of force-perturbation- based criterion as shown in Fig.6.

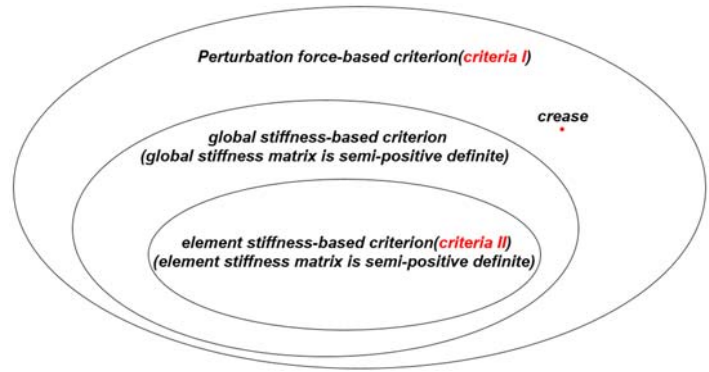

Fig. 6. The relationship between the compliance-based criterion, the global stiffness-based criterion, and the material stiffness-based criterion.

For criterion 1, various perturbation force loadings are needed to identify the instability. To improve the efficiency of process, a small force perturbation can be applied to each node of the finite element model. In this case, some nodes have small displacement, while the relatively low stable nodes have a large displacement, thus the instability phenomenon can easily be captured.

## 3．A stronger criterion for fast determining material instability

Criterion 1 is a universal criterion, which can cover various instability phenomena, but it is a large workload. The positive definiteness of the global stiffness can also be used to determine the occurrence of some instability phenomena, but takes a lot of work to judge whether the global stiffness matrix is positive definite or not. Therefore, for some simple instability phenomena, we want to quickly determine the occurrence of instability through the energy changes of local infinitesimal elements.

Consider a region $V$ of uniform deformation with a deformation gradient $\bar{F}$, as shown in Fig. 6. We investigate whether there exists some perturbation $\delta \boldsymbol{F}$ within a subregion $V_0$ to lower the energy. For a hyperelastic solid, the change in strain energy density $\delta \varphi$ around the equilibrium state $\bar{\boldsymbol{F}}$ can be expressed in terms of a Taylor expansion as

$$\delta \varphi = \left.\frac{\partial \varphi(\boldsymbol{F})}{\partial \boldsymbol{F}}\right|_{\bar{\boldsymbol{F}}} : \delta \boldsymbol{F} + \frac{1}{2} \delta \boldsymbol{F} : \left.\frac{\partial^2 \varphi(\boldsymbol{F})}{\partial \boldsymbol{F} \partial \boldsymbol{F}}\right|_{\bar{\boldsymbol{F}}} : \delta \boldsymbol{F} + O(|\delta \boldsymbol{F}|^3)$$
$$= \boldsymbol{P} : \delta \boldsymbol{F} + \frac{1}{2} \delta \boldsymbol{F} : \left.\frac{\partial \boldsymbol{P}}{\partial \boldsymbol{F}}\right|_{\bar{\boldsymbol{F}}} : \delta \boldsymbol{F} + O(|\delta \boldsymbol{F}|^3) \quad (2)$$

where $\boldsymbol{P}$ is the first Piola-Kirchhoff stress.

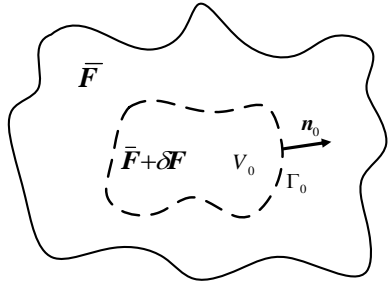

Fig. 7. Schematic of a uniformly deformed solid with a virtual perturbation $\delta \boldsymbol{F}$ inside a subregion $V_0$ bounded by $\Gamma_0$, $\boldsymbol{n}_0$ being the unit normal to $\Gamma_0$.

Since the displacement outside the subregion $V_0$ remains unchanged during the perturbation, the work from external forces is zero, and the global energy change inside the subregion $V_0$ can be expressed as

$$\delta U_{tot} = \int_{V_0} \boldsymbol{P} : \delta \boldsymbol{F} dV + \frac{1}{2} \int_{V_0} \delta \boldsymbol{F} : \left.\frac{\partial \boldsymbol{P}}{\partial \boldsymbol{F}}\right|_{\bar{\boldsymbol{F}}} : \delta \boldsymbol{F} dV + O(|\delta \boldsymbol{F}|^3) \quad (3)$$

The first term in Eq. (3) vanishes due to the following relation:

$$\int_{V_0} \boldsymbol{P} : \delta \boldsymbol{F} dV = \boldsymbol{P} : (\int_{V_0} \delta \boldsymbol{F} dV)$$
$$= \boldsymbol{P} : \int_{V_0} (\delta \boldsymbol{u}) \nabla dV = \boldsymbol{P} : \iint_{\Gamma_0} \delta \boldsymbol{u} \boldsymbol{n}_0 d\Gamma = \boldsymbol{0} \quad (4)$$

where $\boldsymbol{n}_0$ is the unit normal to the boundary $\Gamma_0$ and $\delta \boldsymbol{u}$ is the displacement associated with the perturbation, which is zero at $\Gamma_0$.

Therefore, the energy change in Eq. (3) becomes

$$\delta U_{tot} = \frac{1}{2} \int_{V_0} \delta \boldsymbol{F} : \left.\frac{\partial \boldsymbol{P}}{\partial \boldsymbol{F}}\right|_{\bar{\boldsymbol{F}}} : \delta \boldsymbol{F} dV + O(|\delta \boldsymbol{F}|^3) \tag{5}$$

Note that the deformations near both sides of the instability region interface should satisfy the continuity condition

$$(\bar{\boldsymbol{F}} + \delta \boldsymbol{F}) \cdot \boldsymbol{e}_i = \bar{\boldsymbol{F}} \cdot \boldsymbol{e}_i \tag{6}$$

Here, $\boldsymbol{e}_i$ are noncollinear unit tangent vectors along the interface, where $i = 1, 2$ for three-dimension deformations and $i = 1$ for two-dimension deformations. We refer to a perturbation $\delta \boldsymbol{F}_C$ satisfying Eq. (6) as a compatible perturbation. According to the energy change, we then arrive at the following criterion.

*Element stiffness-based criterion (Criterion 2):*

*Material instability occurs if there exists some compatible deformation perturbation $\delta \boldsymbol{F}_C$ that renders $\delta \boldsymbol{F}_C : \frac{\partial \boldsymbol{P}}{\partial \boldsymbol{F}} : \delta \boldsymbol{F}_C$ negative; otherwise, uniform deformation prevails.*

**Remark 6.** Although the above derivation is based on the assumption of a hyperelastic solid, criterion 2 is also applicable to nonelastic solids, as long as one replaces the change in the strain energy with the change in the work of internal forces, which is given by

$$\delta W_{tot} = \int_{V_0} \boldsymbol{P} : \delta \boldsymbol{F} dV + \frac{1}{2} \int_{V_0} \delta \boldsymbol{F} : \left.\frac{\partial \boldsymbol{P}}{\partial \boldsymbol{F}}\right|_{\bar{\boldsymbol{F}}} : \delta \boldsymbol{F} dV + O(|\delta \boldsymbol{F}|^3) \tag{7}$$

**Remark 7.** The above derivation is based on the Taylor expansion of the energy change with respect to $\delta \boldsymbol{F}$. If an alternative strain measure $\boldsymbol{E}^{(n)}$ (e.g., the Green strain $\boldsymbol{E}^{(1)} = (\boldsymbol{F}^T \cdot \boldsymbol{F} - \boldsymbol{I})/2$) is adopted, the total energy change can be expressed as

$$\delta U_{tot} = \int_{V_0} \boldsymbol{T}^{(n)} : \delta \boldsymbol{E}^{(n)} dV + \frac{1}{2} \int_{V_0} \delta \boldsymbol{E}^{(n)} : \left.\frac{\partial \boldsymbol{T}^{(n)}}{\partial \boldsymbol{E}^{(n)}}\right|_{\bar{\boldsymbol{E}}} : \delta \boldsymbol{E}^{(n)} dV + O(|\delta \boldsymbol{E}^{(n)}|^3) \tag{8}$$

where $T^{(n)}$ is the work-conjugate stress measure of $E^{(n)}$. The term $\int_{V_0} \delta E^{(n)} dV$ is usually nonzero due to the generally nonlinear relation between $\delta E^{(n)}$ and $\delta u$. In this case, the corresponding expression can be obtained via the chain rule as

$$\frac{\partial \boldsymbol{P}}{\partial \boldsymbol{F}} = \frac{\partial \boldsymbol{E}^{(n)}}{\partial \boldsymbol{F}} : \frac{\partial \boldsymbol{T}^{(n)}}{\partial \boldsymbol{E}^{(n)}} : \frac{\partial \boldsymbol{E}^{(n)}}{\partial \boldsymbol{F}} + \boldsymbol{T}^{(n)} : \frac{\partial^2 \boldsymbol{E}^{(n)}}{\partial \boldsymbol{F} \partial \boldsymbol{F}} \tag{9}$$

where the first term determines the material tangent stiffness and the second term the geometric stiffness. For simplicity, we focus on the $\frac{\partial \boldsymbol{P}}{\partial \boldsymbol{F}}$-based criterion.

**Remark 8.** In the preceding derivation, the reference configuration is not limited to a stress-free state, suggesting that the $\frac{\partial \boldsymbol{P}}{\partial \boldsymbol{F}}$-based criterion should remain applicable for any reference configuration. For the same current configuration $\boldsymbol{x}$, we adopt two different reference configurations, $\boldsymbol{X}_{(A)}$ and $\boldsymbol{X}_{(B)}$ to investigate the proposed criterion. The deformation gradients associated with these two reference configurations can be expressed as

$$\boldsymbol{F}_{(A)} = \frac{\partial \boldsymbol{x}}{\partial \boldsymbol{X}_{(A)}}, \quad \boldsymbol{F}_{(B)} = \frac{\partial \boldsymbol{x}}{\partial \boldsymbol{X}_{(B)}} \tag{10}$$

The deformation gradient and Jacobian between the two reference configurations are

$$\boldsymbol{F}_{(BA)} = \frac{\partial \boldsymbol{X}_{(B)}}{\partial \boldsymbol{X}_{(A)}}, \quad J_{(BA)} = \det\left(\frac{\partial \boldsymbol{X}_{(B)}}{\partial \boldsymbol{X}_{(A)}}\right) \tag{11}$$

where $\boldsymbol{F}_{(BA)}$ is a positive definite $3 \times 3$ matrix, so that $J_{(BA)} > 0$.

If the perturbation $\delta \boldsymbol{F}_{(A)}$ in the reference configuration $\boldsymbol{X}_{(A)}$ satisfies compatible condition, i.e., $\delta \boldsymbol{F}_{(A)} \cdot \boldsymbol{e}_{(A)} = \boldsymbol{0}$, the corresponding perturbation $\delta \boldsymbol{F}_{(B)}$ in the reference configuration $\boldsymbol{X}_{(B)}$ should satisfy

$$\delta \boldsymbol{F}_{(B)} \cdot \boldsymbol{e}_{(B)} = \delta \boldsymbol{F}_{(B)} \cdot \boldsymbol{F}_{(BA)} \cdot \boldsymbol{e}_{(A)} = \delta \boldsymbol{F}_{(A)} \cdot \boldsymbol{e}_{(A)} = \boldsymbol{0} \tag{12}$$

i.e., $\delta \boldsymbol{F}_{(B)}$ also satisfies the compatible condition.

The PK-1 stresses in the two reference configurations are related as

$$J_{(BA)} \boldsymbol{P}_{(B)} = \frac{\partial \boldsymbol{X}_{(B)}}{\partial \boldsymbol{X}_{(A)}} \cdot \boldsymbol{P}_{(A)} \tag{13}$$

Therefore, according to Eqs. (10), (11), (12) and (13), under the same compatible perturbation, the energy change under two reference configurations $\boldsymbol{X}_{(A)}$ and $\boldsymbol{X}_{(B)}$ should satisfy the following transformation,

$$\begin{aligned}
\delta \boldsymbol{F}_{(A)} : \frac{\partial \boldsymbol{P}_{(A)}}{\partial \boldsymbol{F}_{(A)}} : \delta \boldsymbol{F}_{(A)} &= \delta \boldsymbol{F}_{(B)} : \boldsymbol{F}_{(BA)} \cdot \frac{\partial \boldsymbol{P}_{(A)}}{\partial \boldsymbol{F}_{(A)}} \cdot \boldsymbol{F}_{(BA)}^{\mathrm{T}} : \delta \boldsymbol{F}_{(B)} \\
&= J_{(BA)} \delta \boldsymbol{F}_{(B)} : \frac{\partial \boldsymbol{P}_{(B)}}{\partial \boldsymbol{F}_{(B)}} : \delta \boldsymbol{F}_{(B)}
\end{aligned} \tag{14}$$

The above relation indicates that the signs of energy change under different reference configurations are consistent because $J_{(BA)} > 0$. If the current configuration is taken as the reference configuration, the results of $\frac{\partial \boldsymbol{P}}{\partial \boldsymbol{F}}$ under initial and current configurations are related as:

$$\frac{\partial \boldsymbol{P}^{(t)}}{\partial \boldsymbol{F}^{(t)}} = J^{-1} \bar{\boldsymbol{F}} \cdot \left( \frac{\partial \boldsymbol{P}^{(0)}}{\partial \boldsymbol{F}^{(0)}} \right) \cdot \bar{\boldsymbol{F}}^T \tag{15}$$

where the superscripts 't' and '0' denote the current and initial configurations, respectively.

**Remark 9.** The onset of material instability cannot be determined by the positive definiteness of either $\frac{\partial \boldsymbol{T}^{(n)}}{\partial \boldsymbol{E}^{(n)}}$ or $\frac{\partial \boldsymbol{P}}{\partial \boldsymbol{F}}$, which might be confusing (Sun and Sacks, 2005). In fact, a material instability can occur under a positive definite $\frac{\partial \boldsymbol{T}^{(n)}}{\partial \boldsymbol{E}^{(n)}}$ due to the geometric stiffness contributed by the stress, such as for the one-dimensional deformations discussed in Section. 3. On the other hand, the material may remain stable for a nonpositive definite $\frac{\partial \boldsymbol{P}}{\partial \boldsymbol{F}}$ (with at least one negative eigenvalue). As an example, $\frac{\partial \boldsymbol{P}}{\partial \boldsymbol{F}}$ will have negative eigenvalues if the sum of the two smallest principal Cauchy stresses is negative (see Appendix A for the proof). On the other hand, the magnitude of the stress could be reduced to a sufficiently low level to ensure material stability.

**Remark 10.** Criterion 2 is used to judge the occurrence of instability based on local energy changes. In fact, it can be proved that criterion 2 is equivalent to the element stiffness matrix is not positive definite in finite element calculation.

Prove:

We can find a *compatible deformation perturbation* $\delta \mathbf{F}$ such that $\delta \mathbf{F} : \frac{\partial \mathbf{P}}{\partial \mathbf{F}} : \delta \mathbf{F} = \delta F_{ij} C^{PF}_{ijkl} \delta F_{kl} < 0$. Crease is a general instability phenomenon, which is independent of the selection of element type. For the convenience of derivation, we take linear triangular element as an example. For triangular element, its element stiffness matrix $\mathbf{K}^{int}$ can be derived from $\frac{\partial \mathbf{P}}{\partial \mathbf{F}} = \mathbf{C}^{PF}$:

$$K^{int}_{iIkK} = \frac{\partial f^{int}_{iI}}{\partial u_{kK}} = \int_{\Omega_0} \frac{\partial N_I}{\partial X_j} C^{PF}_{ijkl} \frac{\partial N_K}{\partial X_l} d\Omega_0 = A_0 \frac{\partial N_I}{\partial X_j} C^{PF}_{ijkl} \frac{\partial N_K}{\partial X_l} \tag{16}$$

We can find a set of displacements $\delta u_{iI}$ such that $\delta u_{iI} \frac{\partial N_I}{\partial X_j} = \partial F_{ij}$, then

$$\begin{aligned} \delta \mathbf{U} : \mathbf{K}^{int} : \delta \mathbf{U} \\ = A_0 \delta u_{iI} \frac{\partial N_I}{\partial X_j} C^{PF}_{ijkl} \frac{\partial N_K}{\partial X_l} \delta u_{kK} \\ = A_0 \delta F_{ij} C^{PF}_{ijkl} \delta F_{kl} < 0 \end{aligned} \tag{17}$$

This proves that criterion 2 is equivalent to the element stiffness matrix is not positive definite. Further, through matlab numerical verification, it is found that for the rigid body constrained element stiffness matrix and the global stiffness matrix composed of the corresponding element stiffness matrix, when the element stiffness matrix is not positive definite, the global stiffness matrix is not positive definite. So criterion 2 is a convenient and stronger condition for judging instability. Therefore, the applicable scope of criterion 1, criterion 2 and the instability criterion based on the global stiffness matrix has the qualitative relationship shown in Fig.6 .

### 3.1 Comparison with existing theories

For necking in a tensile bar, the proposed element stiffness-based criterion gives the following prediction for the onset of necking (Appendix B):

$$\dot{P}_{11}^{(0)} = \left[\frac{\partial \boldsymbol{P}^{(0)}}{\partial \boldsymbol{F}^{(0)}}\right]_{1111} \dot{F}_{11}^{(0)} = 0 \implies \boldsymbol{P}_{cr} = \boldsymbol{P}_{\max}^{(0)}$$

$$\text{or} \quad \dot{P}_{11}^{(t)} = \left[\frac{\partial \boldsymbol{P}^{(t)}}{\partial \boldsymbol{F}^{(t)}}\right]_{1111} \dot{F}_{11}^{(t)} = \left(E^{\sigma T} + \sigma_{11}\right)\dot{F}_{11}^{(t)} = 0 \implies \left(\frac{d\sigma}{d\varepsilon}\right)^{\sigma J} = \sigma \quad (18)$$

which is just the classic Considère criterion (Considère, 1885).

For shear band formation in an elastic-plastic solid, according to Eq. (9) and Rice's assumption $\dot{\boldsymbol{T}} = \boldsymbol{C}^{SE}:\dot{\boldsymbol{E}}$ (namely, the linear comparison solid) (Rice, 1976), $\frac{\partial \boldsymbol{P}}{\partial \boldsymbol{F}}$ can be expressed as

$$\left(\frac{\partial \boldsymbol{P}}{\partial \boldsymbol{F}}\right)_{iJkL} = F_{iI}F_{kK}C_{IJKL}^{SE} + T_{JL}\delta_{ik} \quad (19)$$

which degenerates to the acoustic tensor proposed by Hill and Rice (Hill, 1962; Rice, 1976).

Therefore, the proposed $\frac{\partial \boldsymbol{P}}{\partial \boldsymbol{F}}$-based criterion provides an integrated criterion for two seemingly different classical phenomena (necking and shear banding). In addition, our derivation based on energy analysis is free from complexities associated with various objective stress rates (Bažant et al., 2012).

### 3.2 Strong-anisotropy-induced material instability

After investigating material instability in isotropic materials, we turn our focus to highly anisotropic materials. Kink bands are commonly observed in fiber-reinforced polymers (FRP) subjected to compressive loading along the fiber direction, which is responsible for the relatively low compressive strength compared to the tensile strength (e.g., the compressive strength is 45.3% of the tensile strength for Toray T800S(www.toraycma.com 错误!未找到引用源。)) in FRPs. Kink-band-like instabilities also exist in highly anisotropic layered materials (Ren et al., 2016; Wadee et al., 2004; Guz et al., 2016; Pan et al., 2019).

In contrast to previous studies, which focused on the microstructures and heterogeneity of composites (Budiansky, 1993; Soutis et al., 1995; Kyriakides et al., 1995; Niu et al., 2000; Vogler et al., 2001; Merodio, 2002), in this section, we investigate the compression instability of homogenous solid with strong anisotropy via the proposed criterion 2(see Case II in Fig.8A). To take account of the realistic implications, we assume that the stiffness of Case II is the same as that of layered composite (see Case II in Fig.8A). Case I is composed of two isotropic elastic constituent phases: a hard phase (Young's modulus $E_h$, Poisson's ratio $\nu_h$) and a soft phase (Young's modulus $E_s$, Poisson's ratio $\nu_s$) with volume fractions $c_h$ and $c_s$, respectively. The exact effective stiffness tensor $\mathbf{D}$ of Case I was obtained (Liu et al. 2009). When Case II with the same anisotropy as above is subjected to longitudinal compression, we obtain the critical condition for the kink band via the proposed criterion (Appendix D):

$$\varepsilon_{cr} = \sqrt{1 - \frac{2D_{11}D_{66}}{D_{11}D_{22} - (D_{12})^2 - 2D_{12}D_{66}}} - 1$$

$$\sigma_{cr} = D_{66} \frac{D_{11}D_{22} - (D_{12})^2}{D_{11}D_{22} - (D_{12})^2 - 2D_{12}D_{66}} \sqrt{1 - \frac{2D_{11}D_{66}}{D_{11}D_{22} - (D_{12})^2 - 2D_{12}D_{66}}} \quad (20)$$

If we neglect Poisson's effect and assume that the modulus of the hard phase is much larger than that of the soft phase, our prediction for the critical stress can degenerate into the well-known Rosen's estimate (Rosen, 1965) (Appendix D), i.e.,

$$\sigma_{cr} \to \frac{E_s}{2(1-c_h)}.$$

Therefore, we may predict kink-band-like material instability by considering only the strong anisotropy. In other words, strong anisotropy is a dominant factor for kink bands. Anisotropy can be quantitatively characterized by the energy-ratio-based anisotropy degree (Fang et al., 2019). Fig. 8B shows the variations in the critical strain and normalized compressive strength for kink bands with respect to the reciprocal of anisotropy degree. We can increase the anisotropy degree by increasing the modulus ratio of the hard and soft phases while keeping the other parameters unchanged ($c_h = 0.6$, $v_h = 0.2$, $v_s = 0.35$) (Sun et al., 2017). It can be found that the critical strain decreased with increasing anisotropy degree, indicating early instability even under a small compressive strain. By noting that material instability and an infinite anisotropy degree both imply the loss of positive definiteness of an energy landscape (Fang et al., 2019), one can imagine that for strong anisotropy case, a finite but small perturbation (herein a compressive load) may trigger material instability. When the anisotropy degree is sufficiently large (A>10), the critical strain is almost linear with the reciprocal of the anisotropy degree, and the slope of the asymptote can be determined by the intrinsic properties (Appendix D). Fig. 8B also shows that increasing the modulus of the hard phase alone contributes little to increases in the compressive strength. For example, increasing the modulus ratio $E_h/E_s$ from 50 to 100 increases the compressive strength by only approximately 2% (i.e., $\sigma_{cr}/E_s$ from 0.890 to 0.908), which accounts for the odd similarity in the compressive strength of T800S ($\sigma_{cr}=1.49GPa$, $E_h/E_s=73.5$, Toray(www.toraycma.com)) and T300 ($\sigma_{cr}=1.47GPa$, $E_h/E_s=57.5$, Toray(www.toraycma.com)). Based on the above study, we suggest lowering the anisotropy of composites to avoid a premature kink band, such as improving the modulus of the matrix or adopting weaving arrangement of fibers.

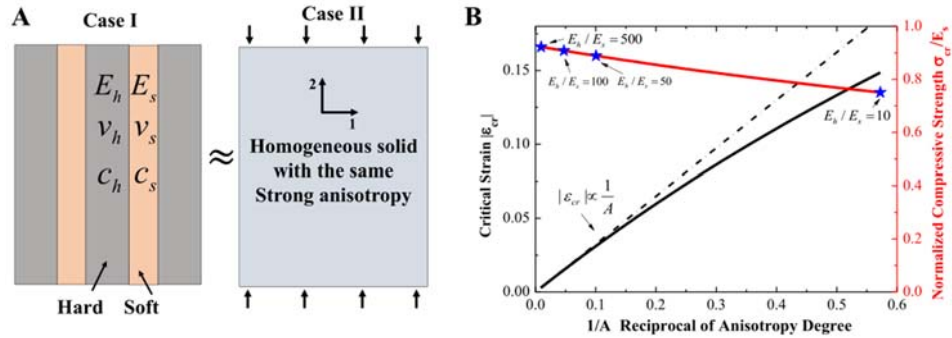

Fig. 8. (A)Schematic of a homogeneous solid under compression (Case II), which possesses the same strong anisotropy as that of layered composite (Case I). (B) Variations in the critical strain and normalized compressive strength for the kink band in Case II with respect to the reciprocal of the anisotropy degree.

## 4. Conclusions

This work is dedicated to establishing an integrated theoretical framework for solid instability with applications to a few elusive instability phenomena. The following conclusions can be drawn:

(1) We have proposed a force-perturbation-based criterion for the onset of solid

instability, which has the widest applicability and can predict instability without priori knowledge of instability configuration.

(2) The crease instability is analyzed in detail, we found that the occurrence of solid instability does not always correspond to the non-positive definiteness of global stiffness matrix. It is because a global stiffness matrix is obtained by smooth perturbation displacement fields, which implicitly require $|\delta \mathbf{u}| \ll 1$, $\left|\dfrac{\partial \delta \mathbf{u}}{\partial \mathbf{x}}\right| \ll 1$. However, when a crease occurs, $\left|\dfrac{\partial \delta \mathbf{u}}{\partial \mathbf{x}}\right| \ll 1$ is not satisfied. But the proposed force-perturbation-based criterion has no limitation on the magnitude of $\left|\dfrac{\partial \delta \mathbf{u}}{\partial \mathbf{x}}\right|$, so it has wider applicability.

(3) An element stiffness-based criterion is proposed in order to fast determine the occurrence of instability. This criterion has been shown to degenerate into the criterion for judging instability of certain known phenomena, such as necking and shear band phenomena. Besides, instability of strongly anisotropic materials is also predicted. For the instability of strongly anisotropic materials, the relationship between the critical strain and the reciprocal of the degree of anisotropy is approximately linear. Increasing the modulus of unidirectional reinforcement does not contribute much to the increase of longitudinal compressive strength.

**Appendix A**

Here we prove that $\frac{\partial \boldsymbol{P}}{\partial \boldsymbol{F}}$ will have negative eigenvalues if the sum of the two smallest principal Cauchy stresses is negative. To illustrate this point, without loss of generality, we take the current configuration as the reference configuration, with coordinate axes coinciding with the principal axes of Cauchy stress. The three principal stresses are denoted as $\sigma_1$, $\sigma_2$, and $\sigma_3$, where $\sigma_1 \leq \sigma_2 \leq \sigma_3$. Considering a hyperelastic material, we can write

$$\frac{\partial \boldsymbol{P}}{\partial \boldsymbol{F}} = \frac{\partial^2 \varphi}{\partial \boldsymbol{F} \partial \boldsymbol{F}} \tag{.1}$$

where $\varphi$ is the energy density function.

We introduce a deformation perturbation in the 1-2 plane as shown in Fig. A1, corresponding to

$$\delta \boldsymbol{F}_\alpha = \begin{bmatrix} 0 & -\alpha & 0 \\ \alpha & 0 & 0 \\ 0 & 0 & 0 \end{bmatrix}, \ \alpha \ll 1 \tag{.2}$$

The associated perturbation in the Green strain is

$$\delta \boldsymbol{E} = \frac{1}{2} \begin{bmatrix} \alpha^2 & 0 & 0 \\ 0 & \alpha^2 & 0 \\ 0 & 0 & 0 \end{bmatrix} \tag{.3}$$

The change in the stored energy can be expressed in terms of the Green strain,

$$\delta \varphi = \frac{\partial \varphi}{\partial \boldsymbol{E}} : \delta \boldsymbol{E} + \frac{1}{2} \delta \boldsymbol{E} : \frac{\partial^2 \varphi}{\partial \boldsymbol{E} \partial \boldsymbol{E}} : \delta \boldsymbol{E} + O\left(|\delta \boldsymbol{E}|^3\right) \tag{.4}$$

Noting that all stress measures should be consistent under the current configuration, the work conjugate relation is given by

$$\frac{\partial \varphi}{\partial \boldsymbol{E}} = \begin{bmatrix} \sigma_1 & 0 & 0 \\ 0 & \sigma_2 & 0 \\ 0 & 0 & \sigma_3 \end{bmatrix} \tag{.5}$$

Substituting Eqs. (A.5) and (A.3) into Eq. (A.4) yields

$$\delta\varphi = \frac{1}{2}(\sigma_1 + \sigma_2)\alpha^2 + O(\alpha^4) \qquad (.6)$$

According to Eq. (A.6), if the sum of the two smallest principal stresses is negative, i.e., $\sigma_1 + \sigma_2 < 0$, nonzero $\alpha$ will lead to negative $\delta\varphi$. This also implies that $\delta\boldsymbol{F}_\alpha : \frac{\partial^2 \varphi}{\partial \boldsymbol{F} \partial \boldsymbol{F}} : \delta\boldsymbol{F}_\alpha$ is negative since $|\delta\boldsymbol{F}_\alpha| \propto \alpha$. Therefore, $\frac{\partial^2 \varphi}{\partial \boldsymbol{F} \partial \boldsymbol{F}}$ possesses at least one negative eigenvalue if $\sigma_1 + \sigma_2 < 0$.

Notably, when $\alpha$ is sufficiently small, the introduced perturbation $\delta\boldsymbol{F}_\alpha$ is close to but not exactly equal to a rigid rotation. Therefore, the corresponding change in the energy density can be nonzero.

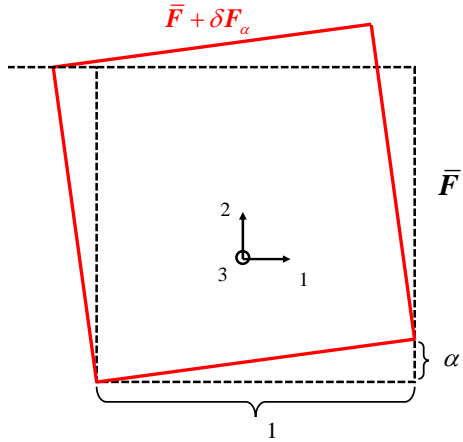

Fig. A1. Schematic of the deformation perturbation applied in the 1-2 plane under the current configuration.

**Appendix B**

Consider a constitutive relation in the form of $\dot{\boldsymbol{T}}^{(1)} = \boldsymbol{C}^{SE} : \dot{\boldsymbol{E}}^{(1)}$, where $\boldsymbol{T}^{(1)}$ is the second Piola-Kirchhoff stress. According to Eqs. (9) and (15), $\frac{\partial \boldsymbol{P}}{\partial \boldsymbol{F}}$ under the current configuration can be expressed as:

$$\frac{\partial \boldsymbol{P}^{(t)}}{\partial \boldsymbol{F}^{(t)}} = J^{-1} \bar{F}_{iI} \bar{F}_{jJ} \bar{F}_{kK} \bar{F}_{lL} C_{IJKL}^{SE} + \sigma_{jl} \delta_{ik} \tag{A.1}$$

The Kirchhoff stress $\boldsymbol{\tau}$, Cauchy stress $\boldsymbol{\sigma}$ and PK-II stress $\boldsymbol{T}^{(1)}$ satisfy the following mutual transformation:

$$\boldsymbol{\tau} = J\boldsymbol{\sigma} = \bar{\boldsymbol{F}} \cdot \boldsymbol{T}^{(1)} \cdot \bar{\boldsymbol{F}}^T \tag{A.2}$$

According to the Lie derivative (or the convective rate) of Kirchhoff stress $\boldsymbol{\tau}$, we have the following relation:

$$\begin{aligned} L_v \boldsymbol{\tau} &= \bar{\boldsymbol{F}} \cdot \dot{\boldsymbol{T}}^{(1)} \cdot \bar{\boldsymbol{F}}^T = \bar{\boldsymbol{F}} \cdot \frac{D}{Dt} \left( \bar{\boldsymbol{F}}^{-1} \cdot \boldsymbol{\tau} \cdot \bar{\boldsymbol{F}}^{-T} \right) \cdot \bar{\boldsymbol{F}}^T \\ &= \dot{\boldsymbol{\tau}} - \boldsymbol{L} \cdot \boldsymbol{\tau} - \boldsymbol{\tau} \cdot \boldsymbol{L}^T = \boldsymbol{\tau}^{\nabla c} \end{aligned} \tag{A.3}$$

where $\boldsymbol{L}$ is the velocity gradient tensor satisfying $\boldsymbol{L} = \dot{\boldsymbol{F}} \cdot \bar{\boldsymbol{F}}^{-1}$. The rate-of-deformation $\boldsymbol{d}$ is defined as the symmetric part of $\boldsymbol{L}$, i.e., $\boldsymbol{d} = (\boldsymbol{L} + \boldsymbol{L}^T)/2$. Therefore, the time derivative of the Green strain $\boldsymbol{E}^{(1)}$ can be expressed as,

$$\dot{\boldsymbol{E}}^{(1)} = \bar{\boldsymbol{F}}^T \cdot \boldsymbol{d} \cdot \bar{\boldsymbol{F}} \tag{A.4}$$

The Truesdell rate of Cauchy stress gives

$$\boldsymbol{\sigma}^{\nabla T} = J^{-1} \cdot \boldsymbol{\tau}^{\nabla c} = \boldsymbol{C}^{\sigma T} : \boldsymbol{d} \tag{A.5}$$

According to Eqs. (B.5), (B.4) and (B.3), we obtain the following relation:

$$C_{ijkl}^{\sigma T} = J^{-1} \bar{F}_{iI} \bar{F}_{jJ} \bar{F}_{kK} \bar{F}_{lL} C_{IJKL}^{SE} \tag{A.6}$$

Under uniaxial tension, the applied force can be related to lateral deformation:

$$\dot{P}_{11}^{(t)} = \left[ \frac{\partial \boldsymbol{P}^{(t)}}{\partial \boldsymbol{F}^{(t)}} \right]_{1111} \dot{F}_{11}^{(t)} = \left( E^{\sigma T} + \sigma_{11} \right) \dot{F}_{11}^{(t)} \tag{A.7}$$

where $E^{\sigma T}$ satisfies the relation $\sigma_{11}^{\nabla T} = E^{\sigma T} d_{11}$.

Assuming transverse isotropy, i.e., $C_{1133}^{\sigma T} = C_{1122}^{\sigma T}$ and $C_{3333}^{\sigma T} = C_{2222}^{\sigma T}$, the lateral deformation can be expressed as follows:

$$d_{22} = d_{33} = -v d_{11}, \quad v = \frac{C_{2211}^{\sigma T}}{C_{2222}^{\sigma T} + C_{2233}^{\sigma T}} \tag{A.8}$$

where $v$ is Poisson's ratio.

According to the Truesdell rate of Cauchy stress $\dot{\boldsymbol{\sigma}} = \boldsymbol{\sigma}^{\nabla T} + \boldsymbol{L} \cdot \boldsymbol{\sigma} + \boldsymbol{\sigma} \cdot \boldsymbol{L}^T - trace(\boldsymbol{d})\boldsymbol{\sigma}$, the uniaxial stress is given by

$$\sigma_{11}^{\nabla T} = E^{\sigma T} d_{11}, \ E^{\sigma T} = C_{1111}^{\sigma T} - 2v C_{1122}^{\sigma T} \\ \dot{\sigma}_{11} = E^{\sigma J} d_{11}, \ E^{\sigma J} = E^{\sigma T} + (1+2v)\sigma_{11} \quad \text{(A.9)}$$

Combining Eqs. (B.9) and (B.7), the critical condition for one-dimensional necking under the current configuration can be expressed as

$$\dot{P}_{11}^{(t)} = \left[\frac{\partial \boldsymbol{P}^{(t)}}{\partial \boldsymbol{F}^{(t)}}\right]_{1111} \dot{F}_{11}^{(t)} = \left(E^{\sigma T} + \sigma_{11}\right)\dot{F}_{11}^{(t)} = \left(E^{\sigma J} - 2v\sigma_{11}\right)\dot{F}_{11}^{(t)} = 0 \quad \text{(A.10)}$$

For an incompressible material, i.e., $v = 0.5$, the corresponding critical condition transforms to

$$E^{\sigma T} + \sigma_{11} = 0 \ \Rightarrow\ E^{\sigma J} = \left(\frac{d\sigma}{d\varepsilon}\right)^{\sigma J} = \sigma_{11} \quad \text{(A.11)}$$

which is the well-known Considère condition for one-dimensional necking.

**Appendix C**

**Compressive instability in a homogeneous solid with strong anisotropy**

The exact effective stiffness tensor $\boldsymbol{D}$ of the layered composite (Fig. 8A in the main text) is obtained based on the following two assumptions (www.toraycma.com): (i) the interfaces are perfectly bonded; (ii) the thickness of each layer is much smaller than the other two dimensions. The explicit components of the effective compliance tensor $\boldsymbol{M}$ under the plane-stress condition are as follows (www.toraycma.com):

$$[\mathbf{M}]_{ij} = \begin{bmatrix} M_{11} & M_{12} & 0 \\ & M_{22} & 0 \\ sym & & M_{66} \end{bmatrix}, \quad \mathbf{D} = \mathbf{M}^{-1}$$

$$M_{11} = \frac{c_h}{E_h} + \frac{c_s}{E_s} - \frac{2c_h c_s (v_h E_s - v_s E_h)^2}{(1-v_h)c_m E_h E_s^2 + (1-v_s)c_h E_s E_h^2},$$

$$M_{22} = \left[ c_h E_h + c_s E_s + \frac{c_h c_s E_h E_s (v_h - v_s)^2}{c_h E_h (1-v_s^2) + c_s E_s (1-v_h^2)} \right]^{-1} \quad \text{(C.1)}$$

$$M_{12} = \frac{c_h v_h + c_s v_s - v_h v_s}{c_h v_s E_h + c_s v_h E_s - c_h E_h - c_s E_s},$$

$$M_{66} = \frac{c_h}{G_h} + \frac{c_s}{G_s}$$

where $c$ denotes the volume fraction, $E$ denotes the Young's modulus, $v$ denotes the Poisson's ratio, $G$ denotes the shear modulus satisfying $G = E/2(1+v)$, and the subscripts $h$ and $s$ denote the hard and soft constituent phases, respectively. We adopt the Green strain $\mathbf{E}$ to take into account the nonlinearity and assume the constitutive relation of Case II obeys $\mathbf{S} = \mathbf{D} : \mathbf{E}$, $\varphi = \frac{1}{2} \mathbf{S} : \mathbf{E}$. Under the given longitudinal compressive deformation $\bar{F}_{22}$, the lateral deformation $\bar{F}_{11}$ for the uniform state can be determined by the following equilibrium condition:

$$\left. \frac{\partial \varphi}{\partial F_{11}} \right|_{\bar{F}_{22}} = 0 \quad \text{(C.2)}$$

Under the specific deformed state ($\bar{F}_{11}$, $\bar{F}_{22}$), the corresponding $\partial \mathbf{P} / \partial \mathbf{F}$ in matrix form can be expressed as

$$\left[\frac{\partial \mathbf{P}}{\partial \mathbf{F}}\right] = \begin{bmatrix} \frac{\partial^2 \varphi}{\partial F_{11} \partial F_{11}} & \frac{\partial^2 \varphi}{\partial F_{11} \partial F_{22}} & \frac{\partial^2 \varphi}{\partial F_{11} \partial F_{12}} & \frac{\partial^2 \varphi}{\partial F_{11} \partial F_{21}} \\ & \frac{\partial^2 \varphi}{\partial F_{22} \partial F_{22}} & \frac{\partial^2 \varphi}{\partial F_{22} \partial F_{12}} & \frac{\partial^2 \varphi}{\partial F_{22} \partial F_{21}} \\ & & \frac{\partial^2 \varphi}{\partial F_{12} \partial F_{12}} & \frac{\partial^2 \varphi}{\partial F_{12} \partial F_{21}} \\ sym & & & \frac{\partial^2 \varphi}{\partial F_{21} \partial F_{21}} \end{bmatrix} \quad (C.3)$$

When $\partial \mathbf{P} / \partial \mathbf{F}$ loses its ellipticity, we obtain the critical condition for the kink band in Case II as follows:

$$\varepsilon_{cr} = F_{22}^{cr} - 1 = \sqrt{1 - \frac{2D_{11}D_{66}}{D_{11}D_{22} - (D_{12})^2 - 2D_{12}D_{66}}} - 1,$$

$$\sigma_{cr} = D_{66} \frac{D_{11}D_{22} - (D_{12})^2}{D_{11}D_{22} - (D_{12})^2 - 2D_{12}D_{66}} \sqrt{1 - \frac{2D_{11}D_{66}}{D_{11}D_{22} - (D_{12})^2 - 2D_{12}D_{66}}} \quad (C.4)$$

(i) Degenerating into Rosen's prediction

If we neglect Poisson's effect, the corresponding stiffness tensor D is as follows:

$$[D]_{ij} = \begin{bmatrix} \left(\frac{c_h}{E_h} + \frac{c_s}{E_s}\right)^{-1} & 0 & 0 \\ 0 & c_h E_h + c_s E_s & 0 \\ 0 & 0 & \left(2\frac{c_h}{E_h} + 2\frac{c_s}{E_s}\right)^{-1} \end{bmatrix} \quad (C.5)$$

According to Eq. (C.4), the corresponding compressive strength can be expressed as

$$\sigma_{cr} = \frac{E_h E_s}{2(c_h E_s + c_s E_h)} \sqrt{1 - \frac{E_h E_s}{(c_h E_s + c_s E_h)^2}} \quad (C.6)$$

When the Young's modulus of the hard phase is much larger than that of the soft phase, our prediction for the compressive strength can degenerate into the well-known Rosen's estimate (6):

$$\sigma_{cr} \to \frac{G_s}{(1-c_h)}, \text{when } E_h \gg E_s \quad (C.7)$$

(ii) Linear relation between the critical strain and reciprocal of the anisotropy degree

When the anisotropy is sufficiently strong (herein $E_h \gg E_s$), the components of stiffness tensor $\mathbf{D}$ would be close to their corresponding asymptotic values, as shown hereafter:

$$\begin{aligned}
D_{22} &\to c_h E_h \\
D_{66} &\to \frac{E_s}{2c_s(1+\nu_s)} \\
D_{11} &\to \frac{E_s}{c_s}\frac{1-\nu_s}{1-\nu_s-2\nu_s^2} \to D_{66}\frac{2(1-\nu_s^2)}{1-\nu_s-\nu_s^2} \\
D_{12} &\to \frac{c_h \nu_h + c_s \nu_s - \nu_h \nu_s}{1-\nu_s} D_{11} \to D_{66}\frac{2(1+\nu_s)(c_h \nu_h + c_s \nu_s - \nu_h \nu_s)}{1-\nu_s-\nu_s^2}
\end{aligned} \quad (C.8)$$

and the asymptotic value of the critical strain can be expressed as

$$\varepsilon_{cr} \to \frac{D_{66}}{D_{22}} \tag{C.9}$$

The anisotropy can be quantified via the following energy-ratio-based anisotropy degree (7):

$$A = \max_{\boldsymbol{\varepsilon}, \mathbf{R}_\varepsilon^{(1)}, \mathbf{R}_\varepsilon^{(2)}} \left\{ \frac{\frac{1}{2}(\mathbf{R}_\varepsilon^{(1)}\boldsymbol{\varepsilon})^T \mathbf{D}(\mathbf{R}_\varepsilon^{(1)}\boldsymbol{\varepsilon})}{\frac{1}{2}(\mathbf{R}_\varepsilon^{(2)}\boldsymbol{\varepsilon})^T \mathbf{D}(\mathbf{R}_\varepsilon^{(2)}\boldsymbol{\varepsilon})} \right\} - 1 \tag{C.10}$$

where $\mathbf{R}_\varepsilon$ is the rotation matrix of strain $\boldsymbol{\varepsilon}$. In this work, the anisotropy degree can be expressed as,

$$A = \frac{D_{11}(\varepsilon_1^{\max})^2 + D_{22}(\varepsilon_2^{\max})^2 + 2D_{12}\varepsilon_1^{\max}\varepsilon_2^{\max} + D_{66}(\gamma_{12}^{\max})^2}{D_{11}(\varepsilon_1^{\min})^2 + D_{22}(\varepsilon_2^{\min})^2 + 2D_{12}\varepsilon_1^{\min}\varepsilon_2^{\min} + D_{66}(\gamma_{12}^{\min})^2} - 1 \tag{C.11}$$

where $\boldsymbol{\varepsilon}^{\max}$ and $\boldsymbol{\varepsilon}^{\min}$ are the normalized strain states ($\|\boldsymbol{\varepsilon}\|=1$) corresponding to the maximum and minimum energy, respectively. As shown in fig. C1, $\boldsymbol{\varepsilon}^{\max}$ and $\boldsymbol{\varepsilon}^{\min}$ converge to a specific state ($\hat{\boldsymbol{\varepsilon}}^{\max}$, $\hat{\boldsymbol{\varepsilon}}^{\min}$) for sufficiently large anisotropy degree. Hence, the corresponding anisotropy degree is,

$$A = \frac{D_{11}(\hat{\varepsilon}_1^{max})^2 + D_{22}(\hat{\varepsilon}_2^{max})^2 + 2D_{12}\hat{\varepsilon}_1^{max}\hat{\varepsilon}_2^{max}}{D_{11}(\hat{\varepsilon}_1^{min})^2 + D_{66}(\hat{\gamma}_{12}^{min})^2} - 1 \quad (C.12)$$

According to Eq. (C.8), $D_{22}$ is much larger than $D_{11}$, $D_{12}$ and $D_{66}$, hence the asymptotic value of the anisotropy degree can be expressed as,

$$A \rightarrow \frac{(\hat{\varepsilon}_2^{max})^2 D_{22}}{\left[(\hat{\gamma}_{12}^{min})^2 + 2(\hat{\varepsilon}_2^{min})^2 \frac{1-v_s^2}{1-v_s-2v_s^2}\right]D_{66}} \quad (C.13)$$

Substituting Eq. (C.9) into (C.13), the critical strain is almost linear with the reciprocal of anisotropy degree when the anisotropy is sufficiently strong ($A > 10$ for this work),

$$\varepsilon_{cr} \rightarrow \frac{(\hat{\varepsilon}_2^{max})^2}{\left[(\hat{\gamma}_{12}^{min})^2 + 2(\hat{\varepsilon}_2^{min})^2 \frac{1-v_s^2}{1-v_s-2v_s^2}\right]} \frac{1}{A} \quad (C.14)$$

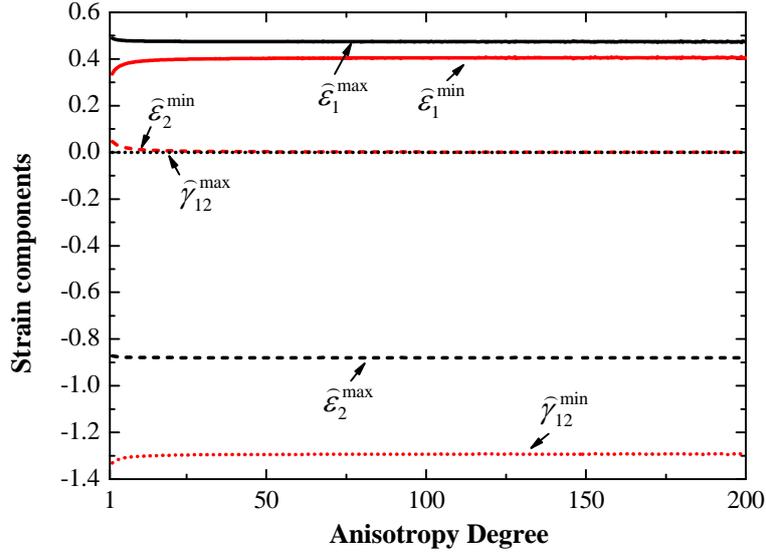

Fig. C1. The variation of normalized strain components with respect to the energy-ratio-based anisotropy degree.

**References**

Audoly, B., Hutchinson, J.W., 2016. Analysis of necking based on a one-dimensional model. J. Mech. Phys. Solids 97, 68–91. https://doi.org/10.1016/j.jmps.2015.12.018

Bazant, Z.P., Belytschko, T.B., 1985. Wave propagation in a strain-softening bar: exact solution. J. Eng. Mech. 111, 381–389. https://doi.org/10.1061/(ASCE)0733-9399(1985)111:3(381)

Bažant, Z.P., Gattu, M., Vorel, J., 2012. Work conjugacy error in commercial finite-element codes: Its magnitude and how to compensate for it. Proc. R. Soc. A Math. Phys. Eng. Sci. 468, 3047–3058. https://doi.org/10.1098/rspa.2012.0167

Belytschko, T., Liu, W., Moran, B., 2000. Nonlinear finite elements for continua and structures, Chichester, New York, John Wiley. https://doi.org/10.1016/S0065-

230X(09)04001-9

Biot, M.A., 1963. Surface instability of rubber in compression. Appl. Sci. Res. Sect. A 12, 168–182. https://doi.org/10.1007/BF03184638

Budiansky B, Fleck NA (1993) Compressive failure of fibre composites. J Mech Phys Solids 41:183-211. doi:10.1016/0022-5096(93)90068-Q.

Cao, Y., Hutchinson, J.W., 2012. From wrinkles to creases in elastomers: the instability and imperfection-sensitivity of wrinkling. Proc. R. Soc. A Math. Phys. Eng. Sci. 468, 94–115. https://doi.org/10.1098/rspa.2011.0384

Chen, D., Jin, L., Suo, Z., Hayward, R.C., 2014. Controlled formation and disappearance of creases. Mater. Horiz. 1, 207-213. https://doi.org/10.1039/C3MH00107E

Ciarletta, P., 2018. Matched asymptotic solution for crease nucleation in soft solids. Nat. Commun. 9, 496. https://doi.org/10.1038/s41467-018-02979-6

Ciarletta, P., Balbi, V., Kuhl, E., 2014. Pattern selection in growing tubular tissues. Phys. Rev. Lett. 113, 248101. https://doi.org/10.1103/PhysRevLett.113.248101

Considère, A., 1885. Mémoire sur l'emploi du fer et de l'acier dans les constructions. Annales des Ponts et Chaussées 6, 574–775.

Crosby, A.J., 2010. Editorial: Why should we care about buckling? Soft Matter 6, 5660. https://doi.org/10.1039/c0sm90040k

De Borst, R., Sluys, L.J., Mühlhaus, H.B., Pamin, J., 1993. Fundamental issues in finite element analyses of localization of deformation. Eng. Comput. 10, 99-121. https://doi.org/10.1108/eb023897

Dervaux, J., Couder, Y., Guedeau-Boudeville, M.A., Ben Amar, M., 2011. Shape transition in artificial tumors: From smooth buckles to singular creases. Phys. Rev. Lett. 107, 018103. https://doi.org/10.1103/PhysRevLett.107.018103

Diab, M., Zhang, T., Zhao, R., Gao, H., Kim, K.-S., 2013. Ruga mechanics of creasing: from instantaneous to setback creases. Proc. R. Soc. A Math. Phys. Eng. Sci. 469, 20120753–20120753. https://doi.org/10.1098/rspa.2012.0753

Fang Y, Wang Y, Imtiaz H, Liu B, Gao H (2019) Energy-Ratio-Based Measure of Elastic Anisotropy. Phys Rev Lett 122:045502. doi:10.1103/PhysRevLett.122.045502.

Flory, P.J., 1979. Molecular theory of rubber elasticity. Polymer (Guildf). 20, 1317–1320. https://doi.org/10.1016/0032-3861(79)90268-4

Fu, Y.B., Ciarletta, P., 2015. Buckling of a coated elastic half-space when the coating and substrate have similar material properties. Proc. R. Soc. A Math. Phys. Eng. Sci. 471, 20140979. https://doi.org/10.1098/rspa.2014.0979

Guz IA, Menshykova M, Soutis C (2016) Internal instability as a possible failure mechanism for layered composites. Philos Trans R Soc A Math Phys Eng Sci 374:20160019. doi:10.1098/rsta.2016.0019.

G'Sell, C., Hiver, J.M., Dahoun, A., 2002. Experimental characterization of deformation damage in solid polymers under tension, and its interrelation with necking. Int. J. Solids Struct. 39, 3857–3872. https://doi.org/10.1016/S0020-7683(02)00184-1

Gent, A.N., Cho, I.S., 1999. Surface Instabilities in Compressed or Bent Rubber

Blocks. Rubber Chem. Technol. 72, 253–262. https://doi.org/10.5254/1.3538798

Ghatak, A., Das, A.L., 2007. Kink instability of a highly deformable elastic cylinder. Phys. Rev. Lett. 99, 76101. https://doi.org/10.1103/PhysRevLett.99.076101

Greer, A.L., Cheng, Y.Q., Ma, E., 2013. Shear bands in metallic glasses. Mater. Sci. Eng. R Reports 74, 71–132. https://doi.org/10.1016/j.mser.2013.04.001

Guo, Y., Ruan, Q., Zhu, S., Wei, Q., Chen, H., Lu, J., Hu, B., Wu, X., Li, Y., Fang, D., 2019. Temperature Rise Associated with Adiabatic Shear Band: Causality Clarified. Phys. Rev. Lett. 122, 015503. https://doi.org/10.1103/PhysRevLett.122.015503

Hadamard, J.S., 1903. Leçons sur la propagation des ondes et les équations de l'hydrodynamique, Hermann, Paris.

Hao, Y., Gong, Z., Xie, Z., Guan, S., Yang, X., Wang, T., Wen, L., 2018. A Soft Bionic Gripper with Variable Effective Length. J. Bionic Eng. 15, 220–235. https://doi.org/10.1007/s42235-018-0017-9

Harren, S. V., Dève, H.E., Asaro, R.J., 1988. Shear band formation in plane strain compression. Acta Metall. 36, 2435–2480. https://doi.org/10.1016/0001-6160(88)90193-9

Hill, R., 1962. Acceleration waves in solids. J. Mech. Phys. Solids 10, 1–16. https://doi.org/10.1016/0022-5096(62)90024-8

Hill, R., Hutchinson, J.W., 1976. Bifurcation Phenomena in the Plane Strain Compression Test. J. Mech. Phys. Solids. 23, 239-264. https://doi.org/10.1016/0022-5096(75)90027-7

Hohlfeld, E., 2013. Coexistence of scale-invariant states in incompressible elastomers. Phys. Rev. Lett. 111, 185701. https://doi.org/10.1103/PhysRevLett.111.185701

Hohlfeld, E., Mahadevan, L., 2012. Scale and nature of sulcification patterns. Phys. Rev. Lett. 109, 025701. https://doi.org/10.1103/PhysRevLett.109.025701

Hohlfeld, E., Mahadevan, L., 2011. Unfolding the Sulcus. Phys. Rev. Lett. 106, 105702. https://doi.org/10.1103/PhysRevLett.106.105702

Holzapfel, G., 2000. Nonlinear solid mechanics: A continuum approach for engineering. New York, John Wiley & Sons. https://doi.org/10.1023/A:1020843529530

Hong, W., Zhao, X., Suo, Z., 2009. Formation of creases on the surfaces of elastomers and gels. Appl. Phys. Lett. 95, 253. https://doi.org/10.1063/1.3211917

Hutchinson, J.W., Tvergaard, V., 1981. Shear band formation in plane strain. Int. J. Solids Struct. 17, 451-470. https://doi.org/10.1016/0020-7683(81)90053-6

Kim, J., Yoon, J., Hayward, R.C., 2010. Dynamic display of biomolecular patterns throughanelastic creasing instability of stimuli-responsive hydrogels. Nat. Mater. 9, 159–164. https://doi.org/10.1038/nmat2606

Koiter, W. T., 1965. On the stability of elastic equilibrium. Ph.D. thesis, Delft University, Holland.

Kyriakides S, Arseculeratne R, Perry EJ, Liechti KM (1995) On the compressive failure of fiber reinforced composites. Int J Solids Struct 32(6–7):689–738. doi:10.1016/0020-7683(94)00157-R.

Liu B, Feng X, Zhang SM (2009) The effective Young's modulus of composites

beyond the Voigt estimation due to the Poisson effect. Compos Sci Technol 69:2198-2204. doi:10.1016/j.compscitech.2009.06.004.

Merodio J, Ogden RW (2002) Material instabilities in fiber-reinforced nonlinearly elastic solids under plane deformation. Arch Mech 54(5-6):525-552.

Mora, S., Abkarian, M., Tabuteau, H., Pomeau, Y., 2011. Surface instability of soft solids under strain. Soft Matter 7, 10612. https://doi.org/10.1039/c1sm06051a

Mullin, T., Deschanel, S., Bertoldi, K., Boyce, M.C., 2007. Pattern transformation triggered by deformation. Phys. Rev. Lett. 99, 084301. https://doi.org/10.1103/PhysRevLett.99.084301

Needleman, A., 1972. A numerical study of necking in circular cylindrical bar. J. Mech. Phys. Solids. 20, 111-127. https://doi.org/10.1016/0022-5096(72)90035-X

Niu K, Talreja R (2000) Modeling of compressive failure in fiber reinforced composites. Int J Solids Struct 37:2405–2428. doi:10.1016/S0020-7683(99)00010-4.

Norris, D.M., Moran, B., Scudder, J.K., Quiñones, D.F., 1978. A computer simulation of the tension test. J. Mech. Phys. Solids. 26, 1-19. https://doi.org/10.1016/0022-5096(78)90010-8

Pan F, et al. (2019) Bending induced interlayer shearing, rippling and kink buckling of multilayered graphene sheets. J Mech Phys Solids 122:340–363. doi:10.1016/j.jmps.2018.09.019.

Ren M, Liu Y, Liu JZ, Wang L, Zheng Q (2016) Anomalous elastic buckling of layered crystalline materials in the absence of structure slenderness. J Mech Phys

Solids 88:83–99. doi:10.1016/j.jmps.2015.12.004.

Rice, J.R., 1976. The localization of plastic deformation. 14th Int. Congr. Theoretical Appl. Mech. 207–220. https://doi.org/10.1.1.160.6740

Rogers, J.A., Someya, T., Huang, Y., 2010. Materials and mechanics for stretchable electronics. Science. 327, 1603-1607. https://doi.org/10.1126/science.1182383

Rosen BW (1965), Mechanics of composite strengthening: Fibre Composite Materials. American Society of Metals, Chapter 3.

Rudykh, S., Boyce, M.C., 2014. Transforming wave propagation in layered media via instability-induced interfacial wrinkling. Phys. Rev. Lett. 112, 034301. https://doi.org/10.1103/PhysRevLett.112.034301

Santisi d'Avila, M.P., Triantafyllidis, N., Wen, G., 2016. Localization of deformation and loss of macroscopic ellipticity in microstructured solids. J. Mech. Phys. Solids 97, 275–298. https://doi.org/10.1016/j.jmps.2016.07.009

Simo, J.., K.S.Pister, 1984. Remarks on rate constitutive equations for finite deformation problems: computational implications. Comput. Methods Appl. Mech. Eng. 46, 201–215. https://doi.org/10.1016/0045-7825(84)90062-8

Soutis C, Turkmen D (1995) Influence of shear properties and fibre imperfections on the compressive behaviour of GFRP laminates. Appl Compos Mater 2(6):327–342. doi:10.1007/BF00564572.

Stören, S., Rice, J.R., 1975. Localized necking in thin sheets. J. Mech. Phys. Solids 23, 421–441. https://doi.org/10.1016/0022-5096(75)90004-6

Sun, W., Sacks, M.S., 2005. Finite element implementation of a generalized Fung-

elastic constitutive model for planar soft tissues. Biomech. Model. Mechanobiol. 4, 190–199. https://doi.org/10.1007/s10237-005-0075-x

Sun W, Guan Z, Li Z, Zhang M, Huang Y (2017) Compressive failure analysis of unidirectional carbon/epoxy composite based on micro-mechanical models. Chinese J Aeronaut 30(6):1907–1918. doi:10.1016/j.cja.2017.10.002.

Tallinen, T., Biggins, J.S., Mahadevan, L., 2013. Surface sulci in squeezed soft solids. Phys. Rev. Lett. 110, 024302. https://doi.org/10.1103/PhysRevLett.110.024302

Tang, S., Gao, B., Zhou, Z., Gu, Q., Guo, T., 2017. Dimension-controlled formation of crease patterns on soft solids. Soft Matter 13, 619–626. https://doi.org/10.1039/c6sm02013e

Triantafyllidis, N., Aifantis, E.C., 1986. A gradient approach to localization of deformation. I. Hyperelastic materials. J. Elast. 16, 225-237. https://doi.org/10.1007/BF00040814

Trivedi, D., Rahn, C.D., Kier, W.M., Walker, I.D., 2008. Soft robotics: Biological inspiration, state of the art, and future research. Appl. Bionics Biomech. 5, 99–117. https://doi.org/10.1080/11762320802557865

Upadhyay, K., Subhash, G., Spearot, D., 2019. Thermodynamics-based stability criteria for constitutive equations of isotropic hyperelastic solids. J. Mech. Phys. Solids 124, 115–142. https://doi.org/10.1016/j.jmps.2018.09.038

Vogler TJ, Hsu S-Y, Kyriakides S (2001) On the initiation and growth of kink bands in fiber composites. Part II: analysis. Int J Solids Struct 38:2653–2682. doi:10.1016/S0020-7683(00)00175-X.

Wadee MA, Hunt GW, Peletier MA (2004) Kink band instability in layered structures. J Mech Phys Solids 52:1071-1091. doi:10.1016/j.jmps.2003.09.026.

Wang, P., Shim, J., Bertoldi, K., 2013. Effects of geometric and material nonlinearities on tunable band gaps and low-frequency directionality of phononic crystals. Phys. Rev. B - Condens. Matter Mater. Phys. 88, 014304. https://doi.org/10.1103/PhysRevB.88.014304

Wang, Q., Zhang, L., Zhao, X., 2011. Creasing to cratering instability in polymers under ultrahigh electric fields. Phys. Rev. Lett. 106, 118301. https://doi.org/10.1103/PhysRevLett.106.118301

Wei, Q., Jia, D., Ramesh, K.T., Ma, E., 2002. Evolution and microstructure of shear bands in nanostructured Fe. Appl. Phys. Lett. 81, 1240–1242. https://doi.org/10.1063/1.1501158

Welker, W., 1990. Why Does Cerebral Cortex Fissure and Fold? Cerebral Cortex, New York, Springer.

Wong, W.H., Guo, T.F., Zhang, Y.W., Cheng, L., 2010. Surface instability maps for soft materials. Soft Matter 6, 5743. https://doi.org/10.1039/c0sm00351d

Yang, Y., Dai, H.H., Xu, F., Potier-Ferry, M., 2018. Pattern Transitions in a Soft Cylindrical Shell. Phys. Rev. Lett. 120, 215503. https://doi.org/10.1103/PhysRevLett.120.215503

Yasnikov, I.S., Vinogradov, A., Estrin, Y., 2014. Revisiting the Considère criterion from the viewpoint of dislocation theory fundamentals. Scr. Mater. 76, 37–40. https://doi.org/10.1016/j.scriptamat.2013.12.009

Zeng, S., Li, R., Freire, S.G., Garbellotto, V.M.M., Huang, E.Y., Smith, A.T., Hu, C., Tait, W.R.T., Bian, Z., Zheng, G., Zhang, D., Sun, L., 2017. Moisture-Responsive Wrinkling Surfaces with Tunable Dynamics. Adv. Mater. 29, 1–7. https://doi.org/10.1002/adma.201700828

Zong, C., Zhao, Y., Ji, H., Han, X., Xie, J., Wang, J., Cao, Y., Jiang, S., Lu, C., 2016. Tuning and Erasing Surface Wrinkles by Reversible Visible-Light-Induced Photoisomerization. Angew. Chemie - Int. Ed. 55, 3931-3935. https://doi.org/10.1002/anie.201510796